%% file: main.tex
\documentclass[10pt,twocolumn,letterpaper]{article}

\usepackage[pagenumbers]{cvpr} 

\input{preamble}

\definecolor{cvprblue}{rgb}{0.21,0.49,0.74}
\usepackage[pagebackref,breaklinks,colorlinks,citecolor=cvprblue]{hyperref}
\usepackage{booktabs}
\usepackage{multirow}
\usepackage[export]{adjustbox}
\usepackage{float}
\def\paperID{14213} 
\def\confName{CVPR}
\def\confYear{2024}

\title{Polyp SAM 2: Advancing Zero shot Polyp Segmentation in Colorectal Cancer Detection}

\author{Mobina Mansoori\\
I-SIP Lab, Concordia University\\
\and
Sajjad  Shahabodini\\
I-SIP Lab, Concordia University
\and
Jamshid Abouei\\
Yazd University\\
\and
Konstantinos N. Plataniotis\\
Multimedia Lab, University of Toronto
\and
Arash Mohammadi\\
I-SIP Lab, Concordia University \\
}

\begin{document}
\maketitle
\input{sec/0_abstract}    
\input{sec/1_intro}
\input{sec/2_ExperimentsandResults}
\input{sec/3_Conclusion}
{
    \small
    \bibliographystyle{ieeenat_fullname}
    \bibliography{main}
}


\end{document}

%% file: preamble.tex
\usepackage[dvipsnames]{xcolor}


%% file: sec/0_abstract.tex

\begin{abstract}
Polyp segmentation plays a crucial role in the early detection and diagnosis of colorectal cancer. However, obtaining accurate segmentation often requires labour-intensive annotations and specialized models. Recently, Meta AI Research released a general Segment Anything Model 2 (SAM 2), which has demonstrated promising performance in several segmentation tasks. In this manuscript, we evaluate the performance of SAM 2 in segmenting polyps under various prompted settings. We hope this report will provide insights to advance the field of polyp segmentation and promote more interesting work in the future. This project is publicly available at \url{https://github.com/sajjad-sh33/Polyp-SAM-2}.
\end{abstract}


%% file: sec/1_intro.tex

\section{Introduction}
Colorectal polyps are common contributors to colorectal cancer, and their early detection is essential for effective treatment. Traditional manual segmentation methods are time-consuming and prone to variability among observers. Recent advances in deep learning have led to automated segmentation techniques, but they often rely on large annotated datasets specific to polyps \cite{li2024asps,jha2024transnetr}.\\
The Segment Anything Model (SAM) \cite{kirillov2023segment} has demonstrated remarkable success in zero-shot image segmentation tasks. SAM’s prompt-based approach allows users to specify objects of interest without the need for additional training. By producing high-quality object masks from input prompts (such as points, boxes, and masks), SAM has garnered attention for various applications. For example, \cite{roy2023sam} assessed SAM’s zero-shot capabilities in organ segmentation tasks. \cite{tang2023can} specifically evaluated SAM’s performance in Camouflaged Object Detection (COD). Meanwhile, \cite{ji2024segment} explored the benefits and limitations of the Segment-Anything Model (SAM) in both computer vision and medical image segmentation tasks. Additionally, \cite{zhou2023can} evaluated the performance of SAM in segmenting polyps, in which SAM is under unprompted settings. Deng et al. \cite{deng2023segment} applied SAM to segment heterogeneous objects in digital pathology.\\
The Segment Anything Model 2 (SAM 2) \cite{ravi2024sam} builds upon the success of its predecessor, SAM, addressing the challenges posed by real time image and video segmentations. Trained on video data, SAM 2 offers real-time capabilities, allowing it to segment entire video sequences based on annotations from a single frame. By leveraging interactions (such as clicks, boxes, or masks) across frames, SAM 2 predicts spatial-temporal masks (referred to as ‘masklets’) for objects. Its unified approach reduces user interaction time and enhances performance, making it a promising tool for various applications, including medical image segmentation and video analysis \cite{yan2024biomedical,shen2024interactive,zhu2024medical, ma2024segment}.\\
In this paper, we explore the application of SAM and SAM 2 to zero-shot polyp image and video segmentation. We evaluate their performance on benchmark datasets and compare them to existing methods. Our results demonstrate the potential of these models for efficient and accurate image and video polyp segmentation, thereby facilitating the way for improved clinical workflows and early cancer detection.\\


%% file: sec/2_ExperimentsandResults.tex

\section{Experiments and Results}

\begin{table*}[hbt!]
\caption{Quantitative Comparison of SAM and SAM 2 on the {\textbf{CVC-ClinicDB}} Dataset.}
\centering
\begin{center}
\fontsize{10pt}{10pt}\selectfont
 \renewcommand{\arraystretch}{1.5}
 \setlength\tabcolsep{15pt}
\begin{tabular}{@{}c|cc|cc|cc|cc@{}}
\hline
\multirow{2}{*}{\textbf{Methods}}& \multicolumn{2}{c|}{\textbf{1 Add - 0 Remove}} & \multicolumn{2}{c|}{\textbf{2 Add - 2 Remove}} & \multicolumn{2}{c|}{\textbf{5 Add - 5 Remove}} & \multicolumn{2}{c}{\textbf{Bounding Box}} \\
\cline{2-9}
                         & mDice           & mIoU           & mDice        & mIoU        & mDice          & mIoU           & mDice       & mIoU        \\
 \cline{1-9} 
SAM  \cite{kirillov2023segment}          & {\textbf{0.626}}          & {\textbf{0.456}}     & {\textbf{0.799}}     & {\textbf{0.665}}     &{\textbf{0.84}}            & {\textbf{0.725}}        & 0.906          & 0.828      \\
SAM 2           & 0.50          & 0.333        & 0.715         & 0.557        & 0.786          & 0.647        &{\textbf{0.93}}   &{\textbf{0.87}}             \\
\hline
\end{tabular}
\label{tab1}
\end{center}
\end{table*}

\begin{table*}[hbt!]
\caption{Quantitative Comparison of SAM and SAM 2 on the {\textbf{Kvasir-SEG}} Dataset.}
\centering
\begin{center}
\fontsize{10pt}{10pt}\selectfont
 \renewcommand{\arraystretch}{1.5}
 \setlength\tabcolsep{15pt}
\begin{tabular}{@{}c|cc|cc|cc|cc@{}}
\hline
\multirow{2}{*}{\textbf{Methods}}& \multicolumn{2}{c|}{\textbf{1 Add - 0 Remove}} & \multicolumn{2}{c|}{\textbf{2 Add - 2 Remove}} & \multicolumn{2}{c|}{\textbf{5 Add - 5 Remove}} & \multicolumn{2}{c}{\textbf{Bounding Box}} \\
\cline{2-9}
                         & mDice           & mIoU           & mDice        & mIoU        & mDice          & mIoU           & mDice       & mIoU        \\
 \cline{1-9} 
SAM  \cite{kirillov2023segment}                        & {\textbf{0.663}}           & {\textbf{0.496}}       & 0.783       & 0.646      &0.837    & 0.72        & 0.855         & 0.747     \\
SAM 2                 & 0.593          & 0.422        & {\textbf{0.809}}         & {\textbf{0.679}}       & {\textbf{0.874}}          & {\textbf{0.771}}       &{\textbf{0.939}}       & {\textbf{0.885}}             \\
\hline
\end{tabular}
\label{tab2}
\end{center}
\end{table*}

\begin{table*}[hbt!]
\caption{Quantitative Comparison of SAM and SAM 2 on the {\textbf{CVC-300}} Dataset.}
\centering
\begin{center}
\fontsize{10pt}{10pt}\selectfont
 \renewcommand{\arraystretch}{1.5}
 \setlength\tabcolsep{15pt}
\begin{tabular}{@{}c|cc|cc|cc|cc@{}}
\hline
\multirow{2}{*}{\textbf{Methods}}& \multicolumn{2}{c|}{\textbf{1 Add - 0 Remove}} & \multicolumn{2}{c|}{\textbf{2 Add - 2 Remove}} & \multicolumn{2}{c|}{\textbf{5 Add - 5 Remove}} & \multicolumn{2}{c}{\textbf{Bounding Box}} \\
\cline{2-9}
                         & mDice           & mIoU           & mDice        & mIoU        & mDice          & mIoU           & mDice       & mIoU        \\
 \cline{1-9} 
SAM  \cite{kirillov2023segment}                     & {\textbf{0.386}}         & {\textbf{0.239}}       & 0.572       & 0.40      &0.649  & 0.481         & {\textbf{0.934}}         & {\textbf{0.876}}       \\
SAM 2           & 0.298         & 0.175        & {\textbf{0.633}}      & {\textbf{0.463}}       & {\textbf{0.689}}        & {\textbf{0.526}}        &0.932       & 0.873          \\
\hline
\end{tabular}
\label{tab3}
\end{center}
\end{table*}

\subsection{Datasets}
To validate the effectiveness of the proposed SAM 2 model, we conducted comparison experiments using six publicly available benchmark colonoscopy datasets. Below, we introduce the details of each dataset. 
1) \textbf{Kvasir{-}SEG} \cite{jha2020kvasir}: This dataset was meticulously curated by the Vestre Viken Health Trust in Norway. It comprises 1,000 polyp images and their corresponding ground truth from colonoscopy video sequences. This dataset is a valuable resource for colonoscopy research. 2) \textbf{CVC-ClinicDB} \cite{bernal2015wm}: This dataset, known as CVC-ClinicDB, was collaboratively curated with the Hospital Clinic of Barcelona, Spain. It comprises 612 images extracted from colonoscopy examination videos, originating from 29 different sequences. 3) \textbf{CVC-ColonDB} \cite{tajbakhsh2015automated}: The dataset comprises 380 polyp images, each accompanied by its corresponding ground truth. Captured at a resolution of $500\times570$, these images were extracted from 15 distinct videos. Experts meticulously ensured that the selected frames represented unique viewpoints by rejecting similar ones. 4) \textbf{ETIS-LaribPolypDB} \cite{silva2014toward}: It comprises 196 polyp images, each captured at a resolution of $966\times1225$ pixels. This dataset plays a crucial role in advancing research related to polyp detection and analysis. 5) \textbf{CVC-300} \cite{vazquez2017benchmark}: The CVC-300 dataset comprises 60 polyp images, each captured at a resolution of $500\times574$ pixels. 6) \textbf{PolypGen} \cite{ali2023multi}: The PolypGen dataset is a comprehensive resource for polyp detection and segmentation. It includes 1,537 polyp images, 2,225 positive video sequences, and 4,275 negative frames. Collected from six different medical centers across Europe and Africa, this dataset provides a diverse set of polyp-related data. Validating the proposed algorithm on the PolypGen dataset enhances the comprehensiveness of the study and brings it closer to real-world scenarios.\\

\begin{table*}[hbt!]
\caption{A quantitative comparison of five public polyp segmentation datasets (\textbf{CVC-ClinicDB, Kvasir, CVC-ColonDB, ETIS, and CVC-300}) with state-of-the-art (SOTA) methods is presented. \textbf{Bold} indicates the best performance.}
\centering
\begin{center}
\fontsize{9pt}{9pt}\selectfont
 \renewcommand{\arraystretch}{1.5}
 \setlength\tabcolsep{10pt}
\begin{tabular}{@{}c|cc|cc|cc|cc|cc@{}}
\hline
\multirow{2}{*}{\textbf{Methods}}& \multicolumn{2}{c|}{\textbf{CVC-ClinicDB}} & \multicolumn{2}{c|}{\textbf{Kvasir-SEG}} & \multicolumn{2}{c|}{\textbf{CVC-ColonDB}} & \multicolumn{2}{c|}{\textbf{ETIS}} & \multicolumn{2}{c}{\textbf{CVC-300}} \\
\cline{2-11}
                         & mDice           & mIoU           & mDice        & mIoU        & mDice          & mIoU           & mDice       & mIoU       & mDice         & mIoU          \\
 \cline{1-11} 
UNet\cite{ronneberger2015u}                     & 0.823           & 0.755          & 0.818        & 0.746       & 0.504          & 0.436          & 0.398       & 0.335      & 0.710         & 0.627         \\
UNet++\cite{zhou2019unet++}                   & 0.794           & 0.729          & 0.821        & 0.744       & 0.482          & 0.408          & 0.401       & 0.344      & 0.707         & 0.624         \\
SFA  \cite{fang2019selective}                    & 0.700           & 0.607          & 0.723        & 0.611       & 0.456          & 0.337          & 0.297       & 0.217      & 0.467         & 0.329         \\
PraNet \cite{fan2020pranet}                  & 0.899           & 0.849          & 0.898        & 0.840       & 0.709          & 0.640          & 0.628       & 0.567      & 0.871         & 0.797         \\
ACSNet  \cite{zhang2020adaptive}                 & 0.882           & 0.826          & 0.898        & 0.838       & 0.716          & 0.649          & 0.578       & 0.509      & 0.863         & 0.787         \\
MSEG   \cite{huang2021hardnet}                  & 0.909           & 0.864          & 0.897        & 0.839       & 0.735          & 0.666          & 0.700       & 0.630      & 0.874         & 0.804         \\
DCRNet   \cite{yin2022duplex}                & 0.869           & 0.800          & 0.846        & 0.772       & 0.661          & 0.576          & 0.509       & 0.432      & 0.753         & 0.670         \\
EU-Net    \cite{patel2021enhanced}               & 0.902           & 0.846          & 0.908        & 0.854       & 0.756          & 0.681          & 0.687       & 0.609      & 0.837         & 0.765         \\
SANet     \cite{wei2021shallow}               & 0.916           & 0.859          & 0.904        & 0.847       & 0.752          & 0.669          & 0.750       & 0.654      & 0.888         & 0.815         \\
MSNet   \cite{zhao2021automatic}                 & 0.918           & 0.869          & 0.905        & 0.849       & 0.747          & 0.668          & 0.720       & 0.650      & 0.862         & 0.796         \\
UACANet    \cite{kim2021uacanet}           & 0.916           & 0.870          & 0.905        & 0.852       & 0.783          & 0.704          & 0.694       & 0.615      & 0.902         & 0.837         \\
C2FNet   \cite{sun2021context}                & 0.919           & 0.872          & 0.886        & 0.831       & 0.724          & 0.650          & 0.699       & 0.624      & 0.874         & 0.801         \\
LDNet    \cite{zhang2022lesion}            & 0.881           & 0.825          & 0.887        & 0.821       & 0.794          & 0.715          & 0.778       & 0.707      & 0.893         & 0.826         \\
SSFormer        \cite{wang2022stepwise}         & 0.906           & 0.855          & 0.917        & 0.864       & 0.802          & 0.721          & 0.796       & 0.720      & 0.895         & 0.827         \\
FAPNet  \cite{zhou2022feature}                 & 0.925           & 0.877          & 0.902        & 0.849       & 0.731          & 0.658          & 0.717       & 0.643      & 0.893         & 0.826         \\
CFA-Net      \cite{zhou2023cross-level}           & 0.933           & 0.883          & 0.915        & 0.861       & 0.743          & 0.665          & 0.732       & 0.655      & 0.893         & 0.827         \\
Polyp-PVT    \cite{dong2021polyppvt}            & 0.948           & 0.905          & 0.917        & 0.864       & 0.808          & 0.727          & 0.787       & 0.706      & 0.900         & 0.833         \\
HSNet     \cite{zhang2022hsnet}               & 0.937           & 0.887          & 0.926        & 0.877       & 0.810          & 0.735          & 0.808       & 0.734      & 0.903         & 0.839         \\
 \cline{1-11} 
SAM-Adapter \cite{chen2023sam}             & 0.774           & 0.673          & 0.847        & 0.763       & 0.671          & 0.568          & 0.590       & 0.476      & 0.815         & 0.725         \\
AutoSAM \cite{hu2023efficiently}                 & 0.751           & 0.642          & 0.784        & 0.675       & 0.535          & 0.418          & 0.402       & 0.308      & 0.829         & 0.739         \\
SAMPath    \cite{zhang2023sam}              & 0.750           & 0.644          & 0.828        & 0.730       & 0.632          & 0.516          & 0.555       & 0.442      & 0.844         & 0.756         \\
SAMed    \cite{li2023polyp-sam}                & 0.404           & 0.273          & 0.459        & 0.300       & 0.199          & 0.115          & 0.212       & 0.126      & 0.332         & 0.202         \\
SAMUS  \cite{lin2023samus}                  & 0.900           & 0.821          & 0.859        & 0.763       & 0.731          & 0.597          & 0.750       & 0.618      & 0.859         & 0.760         \\
SurgicalSAM \cite{yue2024surgicalsam}             & 0.644           & 0.505          & 0.740        & 0.597       & 0.460          & 0.330          & 0.342       & 0.238      & 0.623         & 0.472         \\
MedSAM  \cite{MedSAM}                 & 0.867           & 0.803          & 0.862        & 0.795       & 0.734          & 0.651          & 0.687       & 0.604      & 0.870         & 0.798         \\
Polyp-sam    \cite{ li2024polyp}            & 0.920           & 0.870          & 0.900        & 0.860       & 0.894          & 0.843          & 0.903       & 0.852      & 0.905         & 0.860         \\
ASPS    \cite{li2024asps}                 & 0.951           & 0.906          & 0.920        & 0.858       & 0.799          & 0.701          & 0.861       & 0.769      & 0.919         & 0.852         \\
Polyp-SAM++  \cite{ biswas2023polyp}            & 0.91            & 0.86           & 0.90         & 0.86        & -              & -              & -           & -          & -             & -             \\
$M^{2}$ixNet \cite{zheng2024polyp}                &{\textbf{0.941}}& {\textbf{0.891}}        & 0.929        & 0.881       & 0.820          & 0.855          & 0.891       & 0.866      & 0.895         & 0.861         \\
 \cline{1-11} 
\textbf{SAM 2 (BBox)}        & 0.93           & 0.87         & {\textbf{0.939 }}         &{\textbf{ 0.885}}         & {\textbf{0.934}}           & {\textbf{0.877}}          &{\textbf{0.941}}       & {\textbf{0.89}}      &{\textbf{0.932}}            & {\textbf{0.873}}        \\
\hline
\end{tabular}
\label{tab4}
\end{center}
\end{table*}

\begin{figure*}[hbt!]
  \includegraphics[width=\linewidth]{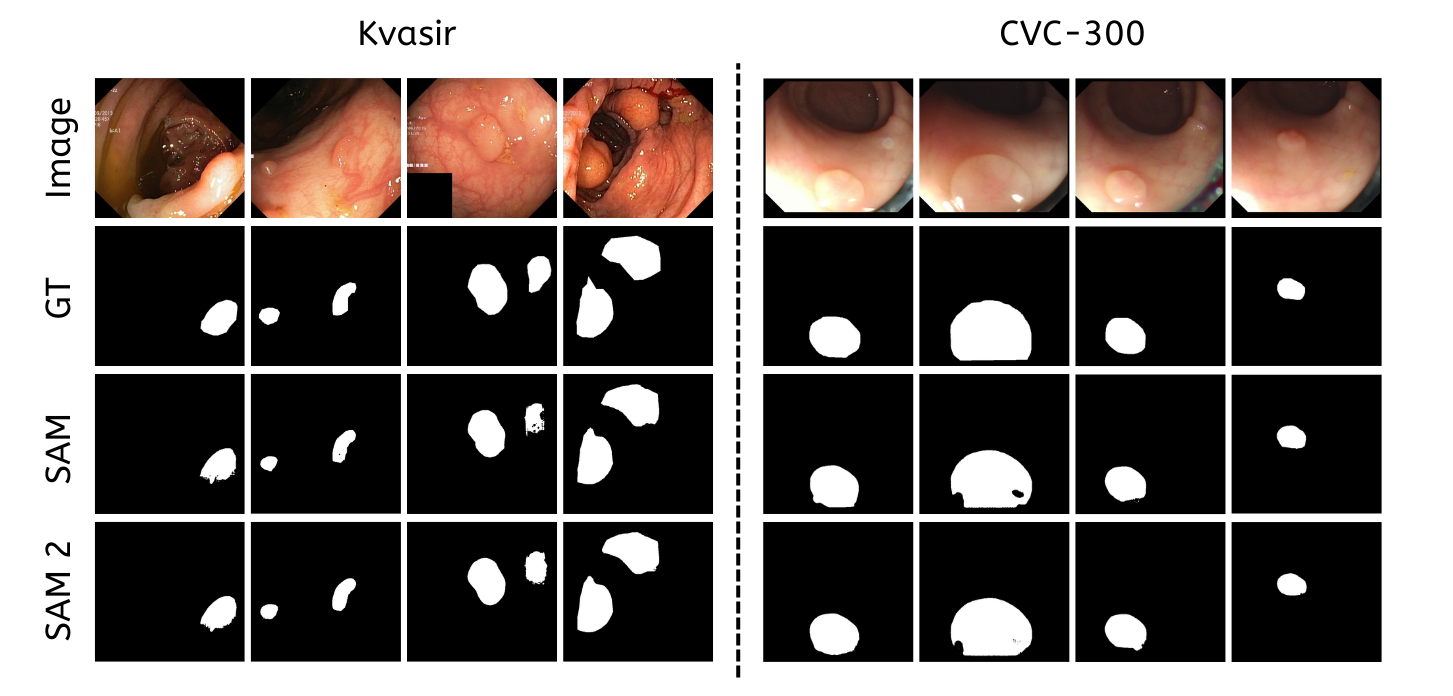}
  \vspace{-20pt} 
  \caption{Qualitative Assessment of Segmentation Outcomes on \textbf{Kvasir-SEG} and \textbf{CVC-300} Datasets using SAM \cite{kirillov2023segment} and SAM 2.}
  \label{Fig1}
  \vspace{-5pt}
  \end{figure*}

\subsection{Evaluation Metrics}
In our experiment,  we employed two widely used metrics for evaluating the effectiveness of  Polyp SAM 2 against other methods of image and video segmentation. Specifically, we evaluated Mean Dice Score (mDice) and Mean Intersection over Union (mIoU).\\

\begin{table}[b!]
\caption{Quantitative results of SAM 2 on PolypGen 23 video sequences. As the input prompt, we are using the bounding box for the first frame.}
\centering
\begin{center}
\fontsize{10pt}{10pt}\selectfont
 \renewcommand{\arraystretch}{1.5}
 \setlength\tabcolsep{20pt}
\begin{tabular}{@{}c|cc@{}}
\hline
\cline{1-3}
  Methods      & mDice           & mIoU                  \\
 \cline{1-3}        
UNet\cite{ronneberger2015u}       &0.4559   &0.4049     \\
UNet++\cite{zhou2019unet++}      &0.4772 &0.4272         \\
ResU-Net++\cite{jha2019resunet++} &0.2105 &0.1589  \\
MSEG   \cite{huang2021hardnet}       & 0.4662 &0.4171    \\
ColonSegNet\cite{jha2021real}       &0.3574 &0.3058      \\
UACANet\cite{kim2021uacanet}        &0.4748  &0.4155      \\
UNeXt\cite{valanarasu2022unext}      &0.2998 & 0.2457         \\
TransNetR\cite{jha2024transnetr}  &0.5168 &0.4717         \\
\cline{1-3} 
\textbf{SAM 2 (BBox)}         &\textbf{0.879}         &\textbf{0.785}         \\
\hline
\end{tabular}
\label{tab5}
\end{center}
\end{table}

\subsection{Quantitative Results between SAM and SAM 2}
First, we compare the zero-shot segmentation results of the SAM and SAM 2 models on the CVC-ClinicDB, Kvasir-SEG, and CVC-300 datasets without fine-tuning. We evaluated four different prompt settings:
\begin{itemize}
\item \textbf{1 Add - 0 Remove}: In this scenario, we provide only one input point to each model. This point is randomly selected from the positive areas (where polyps exist in the image) of the ground truth masks.

\item \textbf{2 Add - 2 Remove}: Here, we give four input points to the model. Two points are randomly chosen from the positive points (where polyps exist) of the ground truth mask, and two from the negative points (where polyps do not exist).

\item \textbf{5 Add - 5 Remove}: In this setting, we provide ten input points—five positive and five negative.

\item \textbf{Bounding Box}: Both the SAM and SAM 2 models receive bounding boxes as input prompts.
\end{itemize}

\cref{tab1,tab2,tab3} show the quantitative comparison of SAM and SAM 2 for these various prompt settings. Results show that increasing the number of input points improves segmentation accuracy for both models. However, bounding box prompts consistently yield better outcomes than point prompts. Overall, SAM 2 performs almost better than the SAM model during bounding box prompt segmentation, highlighting its application to polyp zero-shot segmentation in future works.\\

Furthermore, we delve into the qualitative evaluation of SAM 2 in polyp segmentation tasks. \cref{Fig1} illustrates the visualization results of SAM 2 alongside the SAM model using two selected benchmark datasets. Notably, the SAM 2 model demonstrates superior performance, achieving segmentation results that closely align with the ground truth.

\subsection{Comparison with State-of-the-art Methods}
In \cref{tab4}, we present a quantitative comparison of Polyp SAM 2 with several state-of-the-art methods across five publicly available polyp segmentation datasets mentioned in the 3.1 based on the metrics discussed in 3.2. In particular, we evaluated Polyp SAM 2 against various CNN and ViT models as well as other recent SAM-based segmentation techniques.\\
In our study, CNN-based models include UNet \cite{ronneberger2015u}, UNet++ \cite{zhou2019unet++}, SFA \cite{fang2019selective}, PraNet \cite{fan2020pranet}, ACSNet \cite{zhang2020adaptive}, MSEG \cite{huang2021hardnet}, DCRNet \cite{yin2022duplex}, EU-Net \cite{patel2021enhanced}, SANet \cite{wei2021shallow}, MSNet \cite{zhao2021automatic}, UACANet \cite{kim2021uacanet}, C2FNet \cite{zhou2023cross-level}, LDNet \cite{zhang2022lesion}, FAPNet \cite{zhou2022feature}, and CFA Net \cite{zhou2023cross-level}. For transformer-based models, we evaluated SSFormer \cite{wang2022stepwise}, Polyp PVT \cite{dong2021polyppvt}, and HSNet \cite{zhang2022hsnet}. Furthermore, we explored the impact and effectiveness of SAM-Adapter \cite{chen2023sam}, AutoSAM \cite{hu2023efficiently}, SAMPath  \cite{zhang2023sam}, SAMed \cite{li2023polyp-sam}, SAMUS \cite{lin2023samus}, SurgicalSAM  \cite{yue2024surgicalsam}, MedSAM \cite{MedSAM}, Polyp-SAM \cite{li2024polyp}, ASPS \cite{li2024asps}, Polyp-SAM++ \cite{ biswas2023polyp}, and M2ixNet—all\cite{zheng2024polyp}  of which are built upon the SAM model.\\
Based on the results, we can conclude that SAM 2 is capable of effectively locating and segmenting polyps without additional training. More importantly, among all image segmentation methods, SAM 2 has achieved the highest performance across all scores by a considerable margin (e.g.,1\%, 0.4\% in mDice, mIoU on Kvasir{-}SEG \cite{jha2020kvasir}, 4\%, 2.2\% in mDice, mIoU on CVC-ClinicDB \cite{bernal2015wm}, 5\%, 2.4\% in mDice, mIoU on ETIS-LaribPolypD \cite{silva2014toward} and 1.3\%, 1.2\% in mDice, mIoU on CVC-300 \cite{vazquez2017benchmark} than the second-best methods).

\subsection{Quantitative Results on video polyp segmentation}
In this section, we assess the performance of the SAM 2 model for polyp video segmentation using the PolypGen dataset. As our input prompt, we employ a single bounding box corresponding to the first frame of the video sequence. This bounding box is derived from the ground truth mask of the first frame. Our evaluation results, presented in \cref{tab5}, demonstrate that SAM 2 significantly improves video segmentation performance, achieving a substantial increase of 31.4\% in mean intersection over union (mIoU) compared to the previous state-of-the-art method. Notably, these results were obtained without fine-tuning, distinguishing our approach from prior works.


%% file: sec/3_Conclusion.tex

\section{Conclusion}
 In this study, we investigated the effectiveness of two zero-shot segmentation models: SAM and SAM 2. Our evaluation focused on medical image segmentation tasks, specifically polyp image and video segmentation. SAM 2 consistently outperformed SAM across various metrics, achieving significant improvements without fine-tuning. Bounding box prompts yielded better outcomes for both models, highlighting SAM 2’s practical applicability. Furthermore, SAM 2 surpassed state-of-the-art methods, showcasing its potential for clinical applications. Future research could explore fine-tuning strategies and generalization to other medical imaging tasks.
